\documentclass[aps,showpacs,preprint]{revtex4}
\usepackage{graphicx}
\begin{document}
\title{Detection of quantum critical points by a probe qubit
\footnote{Corresponding authors:
Jingfu Zhang, zhangjfu2000@yahoo.com;\\
Dieter Suter, Dieter.Suter@uni-dortmund.de}}
\author{
Jingfu Zhang, Xinhua Peng,  Nageswaran Rajendran, and Dieter Suter}
\affiliation{Technische Universit\"{a}t Dortmund, 44221 Dortmund, Germany\\
}
\date{\today}

\begin{abstract}

Quantum phase transitions occur when the ground state of a quantum
system undergoes a qualitative change when an external control
parameter reaches a critical value.
Here, we demonstrate a technique for studying quantum systems undergoing
a phase transition by coupling the system to a probe qubit.
It uses directly the increased sensibility of the quantum system
to perturbations when it is close to a critical point.
Using an NMR quantum simulator, we demonstrate this measurement
technique for two different types of quantum phase transitions in an Ising
spin chain.
\end{abstract}

\maketitle

{\it Introduction.---} Phase transitions describe sudden changes in
the properties of a physical system when an external control
parameter changes through some critical value. If the system under
consideration is a quantum mechanical system in its ground state,
i.e. at zero temperature, and the phase transition occurs as a
function of a non-thermal control parameter, we speak of quantum
phase transitions (QPTs) ~\cite{QPTbook}. Examples include the
transitions in superconductors ~\cite{BCS} and fractional quantum
Hall systems ~\cite{HallEffect}. Related phenomena have also been
experimentally observed in heavy Fermion systems ~\cite{Custers},
common metals ~\cite{Yeh}, and in Bose-Einstein condensates
~\cite{Greiner02}. QPTs occur as a result of competing interactions
and the different phases often show different types of correlations
between the constituents, with correlation lengths that can become
arbitrarily large. When specific quantum effects of phase
transitions are of interest, it is therefore natural to compare
entanglement in the different phases ~\cite{entangedQPT,Vidal03}.

Experimental observations of QPTs are relatively straightforward
when they are accompanied by a change of a suitable order parameter,
such as the conductivity or susceptibility in superconductors or the
total magnetization in some spin chains \cite{entangedQPT,Peng05}.
However, such global measurements cannot provide all the details
and they are not suitable for closer investigations of
the systems in the interesting area close to the critical points.
Moreover, not all order parameters can be measured by global
measurements. A complete analysis of the system is provided by
quantum state tomography \cite{Chuang,Peng05}, but this approach
scales very poorly with the size of the system.

As a possible alternative for closer investigations of quantum
systems in the vicinity of critical points, it was suggested to
compare the evolution of systems at slightly different values of the
control parameter. This approach may be considered as a
visualization of "quantum fluctuations". Different possibilities
exist for comparing these evolutions, some of which have been called
Loschmidt echo (LE) or fidelity decay ~\cite{LEReview}. In the
vicinity of critical points, the systems are expected to be much
more susceptible to external perturbations than in the center of a
phase ~\cite{sunPRL06}. Such a comparison is possible by coupling
the system under study to a second quantum system, consisting in the
simplest case of a single qubit. The two states of the probe qubit
can then be used to probe the system under two different values of
the control parameter. The signal obtained in this case corresponds
to the overlap of two states evolving under slightly different
control parameters.

In this Letter we implement this protocol in a nuclear magnetic
resonance (NMR) quantum information processor. The system undergoing
the QPT corresponds to an Ising-type spin chain and the control
parameter to a longitudinal magnetic field. In a purely longitudinal
field, the ground state is degenerate at the critical points. This
degeneracy is lifted if the magnetic field contains a transverse
component. For the longitudinal as well as for the transverse case,
we measure the QPT by coupling the spin chain to a probe qubit.

\begin{figure}
\includegraphics[width=5in]{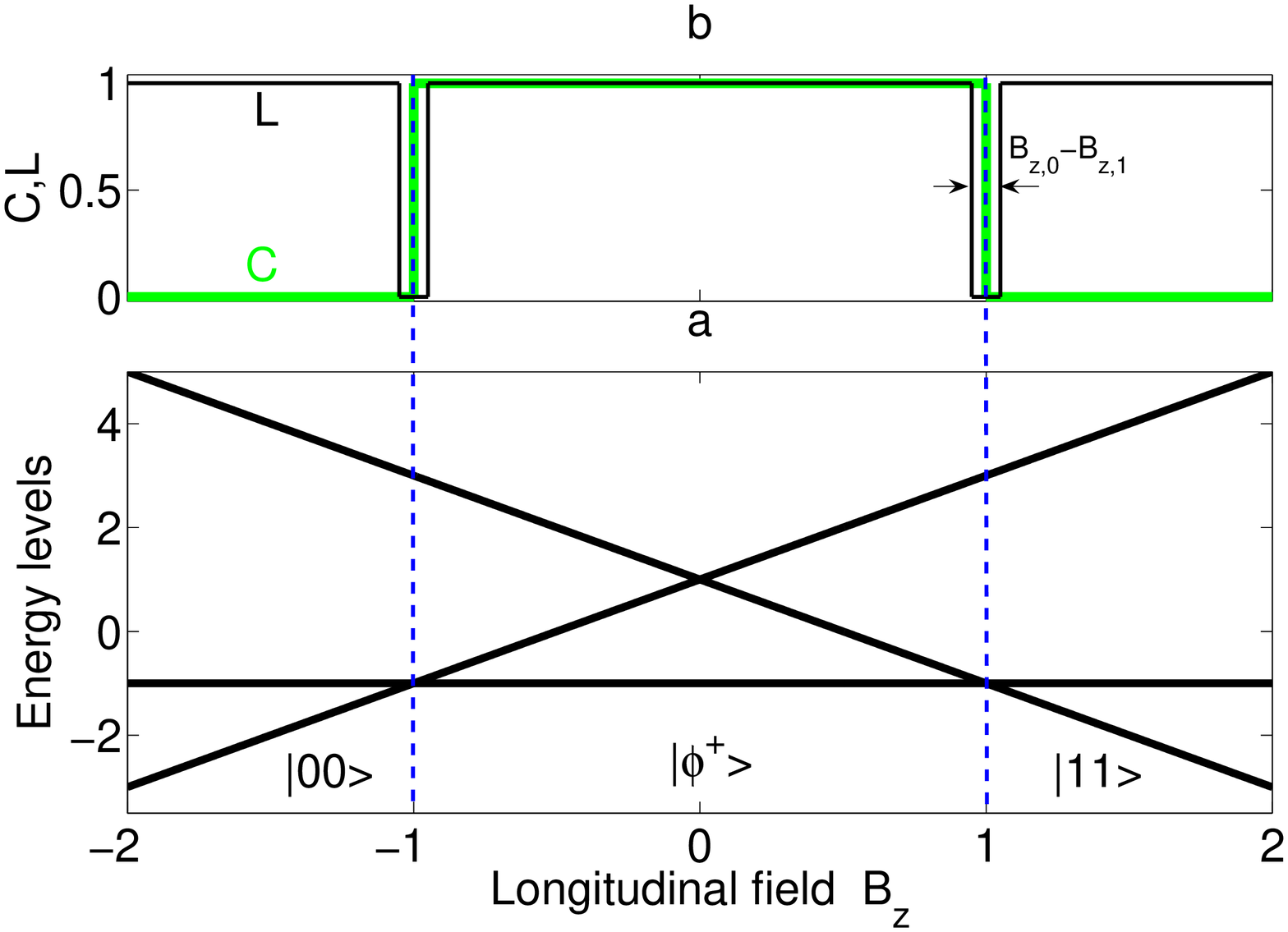}
\caption{(Color online) (a) The energy levels of the system. (b)
Concurrence (thick line) and overlap $L$ (thin) of the ground
state.} \label{con}
\end{figure}

{\it Level-crossing.---}
 We first consider a QPT in the Ising model in a minimal system
consisting of two spins 1/2.
Its Hamiltonian is
\begin{equation}\label{hamE}
    H^{s}=\sigma^{1}_{z}\sigma^{2}_{z}
    +B_{z}(\sigma^{1}_{z}+\sigma^{2}_{z}),
\end{equation}
where the $\sigma^{i}_z$ are Pauli operators and $B_z$ is a magnetic field.
The units have been chosen such that the coupling constant between the
two qubits is 1.
For the purpose of this paper, it is sufficient to consider the triplet manyfold.
Within this subsystem, the ground state depends on the field strength:
\begin{equation}\label{ground}
    |\psi_{g}(B_z)\rangle=\left\{\begin{array}{cc}
    |00\rangle & (B_z\leq-1)\\
    |\phi^{+}\rangle& (-1\leq B_z\leq 1)\\
     |11\rangle & (B_z\geq 1) ,
   \end{array} \right.
\end{equation}
where $|\phi^{+}\rangle=(|01\rangle +|10\rangle)/\sqrt{2}$.
Figure \ref{con} shows the energy levels of the system and the concurrence
of the ground state.
Obviously $B_{z}=\pm1$ are the critical points.
The low-field phase is maximally entangled ($C=1$), while the high-field phases
correspond to product states ($C=0$).

In this system, the QPTs occur at points where the ground state is
degenerate. Close to this critical point, it is therefore very
susceptible to small perturbations. If we couple it to a probe qubit
(which we label 0) via the interaction $\varepsilon
\sigma_z^0(\sigma_z^1 + \sigma_z^2)$, the total (three-qubit) system
can be decomposed into two subsystems, in which qubits 1 and 2 ``see"
an effective field $B_z \pm \varepsilon$.
If these two fields fall
on different sides of the critical point, the ``state overlap" ~\cite{zanardi:031123}
$L=|\langle \psi_{g}(B_{z,0})|\psi_{g}(B_{z,1})\rangle|^{2}$
vanishes, otherwise it is unity, as shown by the thin
line in Figure \ref{con}.
Here, $B_{z,0} = B_z + \varepsilon$
specifies the effective field for the subsystem coupled to $\vert 0
\rangle_0$ and correspondingly for the other subsystem.
In the extreme case where the two states of the probe qubit are orthogonal
($L=0$), the probe qubit has ``measured" the quantum system \cite{Leg02}.

\begin{figure}
\includegraphics[width=5in]{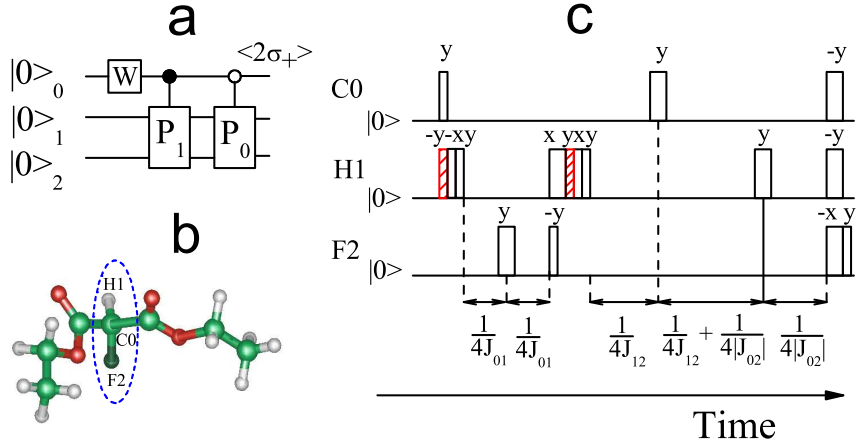}
\caption{(a) Quantum network for measuring $L$. $W$ denotes the
Walsh-Hadamard transform, and $\sigma_{+}=(\sigma_{x}+i
\sigma_{y})/2$. The controlled operations $P_{0}$ and $P_{1}$ denote
the evolutions for preparing $|\psi_{g}(B_{z,0})\rangle$ and
$|\psi_{g}(B_{z,1})\rangle$, if qubit 0 is in state $|0\rangle$ or
$|1\rangle$, respectively. (b) Chemical structure of Diethyl
fluoromalonate. The three qubits are marked by the dashed oval. (c)
Pulse sequence for measuring the state overlap. The narrow unfilled
rectangles denote $\pi/2$ pulses, and the wide ones denote $\pi$
pulses. The striped rectangles denote $\pi/4$ pulses. The directions
along which the pulses are applied are denoted by $\pm x$ and $\pm
y$. The durations of the pulses are so short that they can be
ignored.} \label{sta_Net}
\end{figure}

   To measure $L$, we first initialize the system and probe qubits into
the ground state $\vert 000 \rangle$. From there a Walsh-Hadamard
transform places the probe qubit into the symmetric superposition
state $\vert +\rangle = (\vert 0 \rangle + \vert 1
\rangle)/\sqrt{2}$. We then use the interaction between the probe
and the system to apply a conditional evolution to the two system
qubits 1 and 2: if the probe qubit is in state $\vert 0 \rangle$,
the system evolves from $\vert 00 \rangle \rightarrow
|\psi_{g}(B_{z,0})\rangle$, and if qubit 0 is in state $\vert 1
\rangle$, the system evolves from $\vert 00 \rangle \rightarrow
|\psi_{g}(B_{z,1})\rangle$. The network representation of this
process is shown in Figure \ref{sta_Net}(a) \cite{Somma02}. $P_{0}$
and $P_{1}$ denote the conditional evolutions. The output of the
network is
\begin{equation}\label{finals}
   |\Psi\rangle=[|0\rangle|\psi_{g}(B_{z,0})\rangle+
|1\rangle|\psi_{g}(B_{z,1})\rangle]/\sqrt{2}.
\end{equation}
Taking the trace over the (12)-system, one obtains the reduced
density matrix $\rho^{(0)}$ of the probe qubit.
The off-diagonal
elements are $\rho^{(0)}_{12} = \rho^{(0)\dagger}_{21} =
\langle\psi_{g}(B_{z,0})|\psi_{g}(B_{z,1})\rangle$. Hence the
overlap $L$ can be obtained by measuring
$L=4|\langle \sigma_{+}^0 \rangle|^{2}$,
the transverse magnetization of the probe qubit,
which can be observed as a free induction decay.

  For the experimental implementation, we chose the nuclear spins of
$^{13}$C, $^{1}$H, and $^{19}$F of Diethyl-fluoromalonate as qubits,
shown in Figure \ref{sta_Net}(b).
The scalar coupling constants are $J_{12}=47.6$ Hz, $J_{10}=161.3$
Hz and $J_{20}=-192.2$ Hz.
   The sample consisted of a 2.3:1 mixture of unlabeled  Diethyl fluoromalonate
 and d6-acetone. Molecules
with a $^{13}$C nucleus, which we used as the quantum
register, were therefore present at a concentration of about $0.7
\%$.

\begin{figure}
\includegraphics[width=5in]{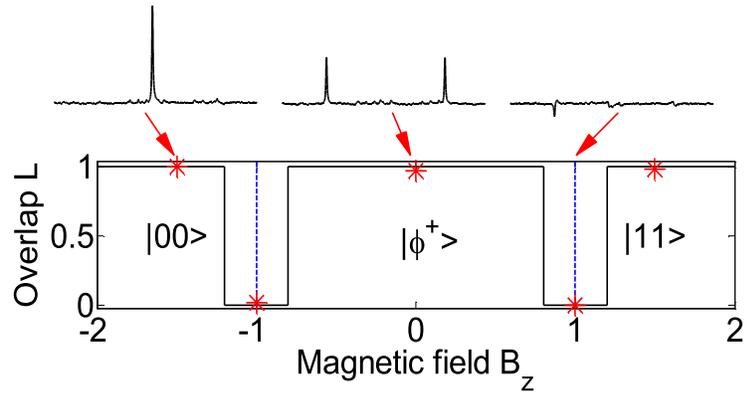}    
\caption{Theoretical (line) and measured (asterisks) overlap $L$ for
the level-crossing case. Three NMR spectra illustrate the signals
corresponding to $B_z=-1.5$, $0$, and $+1$, respectively. }
\label{sta_exp}
\end{figure}

The effective pure state $|000\rangle$ was prepared by spatial
averaging ~\cite{pseudopure1}. We implemented the quantum network of
Figure \ref{sta_Net}(a) for five cases corresponding to $B_z=-1.5$,
$-1$, $0$, $1$, $1.5$,
 respectively.
As an example, Figure \ref{sta_Net}(c) shows the pulse sequence when
$B_z=-1$. Figure \ref{sta_exp} shows the experimental results. The
spectra on top were measured for the values of $B_z$ = -1.5, 0, and
+1, and the asterisks indicate the integrated signal amplitudes.
Clearly, the integrated signal essentially vanishes at the quantum
critical point, while it remains close to the maximum inside the
three phases, in excellent agreement with the theoretical
expectation.

{\it Avoided level-crossing.---} If we add a transverse field to the
system, the QPTs are no longer singular points, but they acquire a
finite width. The modified Hamiltonian is
\begin{equation}\label{hamET}
    H^{s}_T = \sigma^{1}_{z}\sigma^{2}_{z}+B_{x}(\sigma^{1}_{x}+\sigma^{2}_{x})
    +B_{z}(\sigma^{1}_{z}+\sigma^{2}_{z}) .
\end{equation}
As long as $B_x\ll 1$, the ground states are very close to those of
Eq. (\ref{ground}), except in the vicinity of the critical points, where the
transverse field mixes them, thus avoiding the level crossing. In
this region, it is sufficient to consider the two lowest energy
states. They form a two-level system that can be described by the
effective Hamiltonian
$$
H_\mathit{eff} = - |B_z| I + (1 - |B_z|) \sigma_z + \sqrt{2} B_x
\sigma_x,
$$
where $I$ denotes the unit operator.
Within this approximation, the ground state is
$\vert \psi_g(B_x,B_z) \rangle = \vert ll \rangle \cos (\varphi/2) +
\vert \phi^{+} \rangle \sin (\varphi/2)$, 
where
$\tan\varphi = \sqrt{2} B_x/(1-|B_z|) . \label{e.phi}$
When $B_z > 0$, $\vert ll \rangle = \vert 11 \rangle$; when $B_z <
0$, $\vert ll \rangle = \vert 00 \rangle$.

The coupling of this system to a probe qubit provides us with a natural way
of measuring the phase transition.
As before, we use an Ising-type interaction, which results in the total Hamiltonian
\begin{equation}\label{gham}
    H=H^{s}_T+\varepsilon\sigma^{0}_{z}(\sigma^{1}_{z}+\sigma^{2}_{z}).
\end{equation}

To measure the QPT, we first initialize the system into the ground
state $ \vert \psi_{g}(B_x,B_z)\rangle$ in a given longitudinal
field $B_z$. As we turn on the coupling to the probe qubit
initially in $\vert +\rangle$, the combined system splits into two
subsystems, corresponding to the two eigenstates of the probe
qubit. In the two subsystems, the effective longitudinal field
acting on qubits 1 and 2 is $1-|B_z| \pm \varepsilon$, i.e. the
coupling shifts the two subsystems in opposite directions along
the $B_z$-axis. The eigenstates of the two subsystems are
therefore also different. In terms of the mixing angle $\varphi$,
the sensitivity of the basis states to the variation of the
longitudinal field can be quantified as
\begin{equation}\label{sensitive}
d \varphi/d |B_z| =\sqrt{2} B_x/[2 B_x^2 + (1-|B_z|)^2] .
\end{equation}
Apparently, this is a resonant effect: The sensitivity reaches a
maximum at the QPT and falls off with the distance from the critical
points like a Lorentzian. The full width of this "resonance line" is
equal to the splitting $2\sqrt{2}B_x$ of the two lowest energy
levels at the critical point.

To measure this behavior, we initialize the probe qubit into the
 $\vert +\rangle$ state.
 As we turn on the coupling to the
system, each subsystem is no longer in an eigenstate, but starts to
evolve in its new basis.
The initial state $ \vert +\rangle\vert
\psi_{g}(B_x,B_z)\rangle$ evolves as
\begin{equation}\label{final}
   |\Psi(\tau)\rangle =
   [|0\rangle \vert \Psi_{0}(\tau)\rangle +
   |1\rangle|\Psi_{1}(\tau)\rangle]/\sqrt{2}.
\end{equation}
Here $\vert \Psi_{0}(\tau)\rangle$ describes the two system qubits
coupled to the state $\vert 0 \rangle$ of the probe qubit, evolving
under the Hamiltonian
$H_{0}=H^{s}_T+\varepsilon(\sigma^{1}_{z}+\sigma^{2}_{z})$.
Similarly, $\vert \Psi_{1}(\tau)\rangle$ describes the probe qubit
in state $\vert 1 \rangle$ and evolves under
$H_{1}=H^{s}_T-\varepsilon(\sigma^{1}_{z}+\sigma^{2}_{z})$. We use
this differential evolution for measuring the QPT via the overlap
$L=|\langle\Psi_{0}(\tau)|\Psi_{1}(\tau)\rangle|^{2}$.

\begin{figure}
\includegraphics[width=5in]{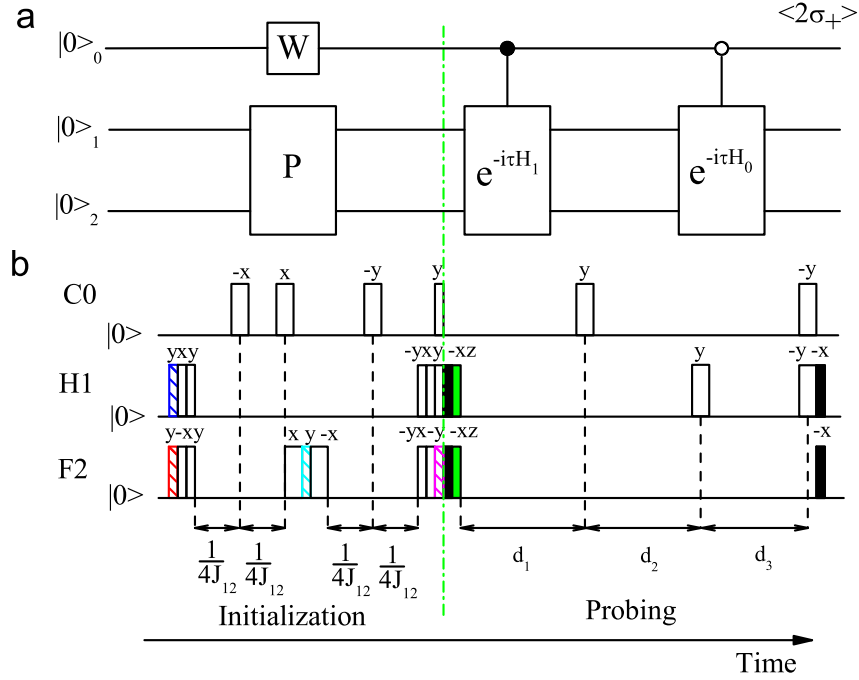}
\caption{ (Color online) Quantum circuit (a) and pulse sequence (b)
for measuring the overlap $L$ for the avoided level-crossing case.
$P$ denotes the preparation of $|+\rangle|\psi_{g}(B_x,B_z)\rangle$
from $|000\rangle$. In the initialization section, the width of the
filled pulse applied to H1 is $\theta-\frac{\pi}{2}$, and the width
of the filled pulses applied to F2, from left to right, are
$(\pi-\alpha)/2$, $(\alpha-\beta)/2$, and $(\pi-\beta)/2$,
respectively, with $\tan(\alpha/2)=-\sqrt{2}c_1/c_+$,
$\tan(\beta/2)=-c_+/\sqrt{2}c_0$,
$\tan(\theta/2)=\sin(\beta/2)/\cos(\alpha/2)$. $c_0$, $c_+$ and
$c_1$ denote the amplitudes of $|00\rangle$, $|\phi^{+}\rangle$ and
$|11\rangle$ in $|\psi_{g}(B_x,B_z)\rangle$. In the probing section,
the rectangles filled by heavy and light color denote the pulses
with width $\tau B_x$ and $2\tau B_z$, respectively, and
$d_1=\frac{\tau}{\pi}(\frac{\varepsilon}{J_{01}}+\frac{1}{J_{12}})$,
$d_2=\frac{\tau}{\pi}(\frac{\varepsilon}{|J_{02}|}+\frac{1}{J_{12}})$,
and
$d_3=\frac{\tau\varepsilon}{\pi}(\frac{1}{J_{01}}+\frac{1}{|J_{02}|})$.}
\label{network}
\end{figure}

Figure \ref{network} shows the quantum circuit and pulse sequence
that implement this measurement. The initialization section prepares
$|+\rangle|\psi_{g}(B_x,B_z)\rangle$, and the probing section
implements the global evolution $U(\tau)$ approximately by
decomposing it into
$e^{-i\tau B_x(\sigma_{x}^{1}+\sigma_{x}^{2})/2}$$e^{-i\tau
\sigma_{z}^{1}\sigma_{z}^{2}}$$e^{-i\tau \varepsilon
\sigma_{z}^{0}\sigma_{z}^{1}}
$$e^{-i\tau\varepsilon \sigma_{z}^{0}\sigma_{z}^{2}}e^{-i\tau
B_z(\sigma_{z}^{1}+\sigma_{z}^{2})}$$e^{-i\tau
B_x(\sigma_{x}^{1}+\sigma_{x}^{2})/2}$. 
In the experiments, we measured the overlap $L$ for a transverse
field strength of $B_{x}=0.1$, coupling strengths of
$\varepsilon=0.2$ and $0.3$, and a range of longitudinal fields,
$-2\leq B_z\leq2$. The evolution time was set to $\tau=1.6$. The
approximations used in the implementation of $U(\tau)$ reduce
the fidelity by less than 1.4\%. 

\begin{figure}
\includegraphics[width=5in]{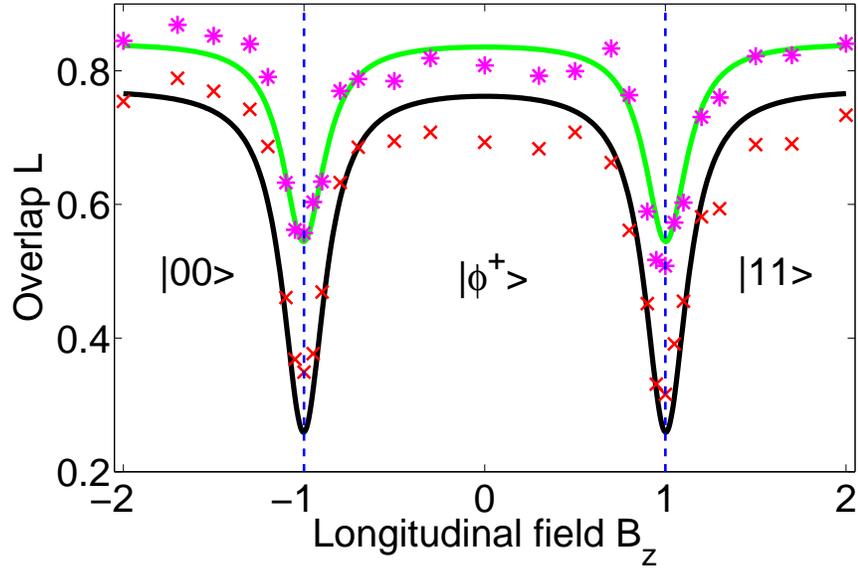}
\caption{(Color online) Experimental overlap $L$ for
$\varepsilon=0.2$ marked by "*" and $\varepsilon=0.3$ marked by
"$\times$". The experimental data are fitted to $aL_0$, and yielded
$a = 0.84$ and $0.77$, respectively, shown as the dark and light
curves, where $L_0$ denotes the corresponding theoretical result. }
\label{Res}
\end{figure}

  The experimental results are shown as Figure
\ref{Res}, where the experimentally measured overlaps $L$ are marked
by "*" and "$\times$" for $\varepsilon=0.2$ and $0.3$, respectively.
The experimental data are fitted to $aL_0$, where $L_0$ denotes the
corresponding theoretical result. The best agreement was obtained
for $a = 0.84$ and $0.77$, respectively; the corresponding functions
are shown as the dark and light curves. Obviously the critical
points are correctly identified by the minima of $L$, indicating
increased sensitivity of the ground state to the perturbation by the
probe qubit. The differences between the theoretical and
experimental values are mainly caused by imperfections of the radio
frequency pulses, inhomogeneities of magnetic fields and
decoherence.

   {\it Discussion and Conclusion.---}   In conclusion, we have shown that a probe qubit
 can be used to detect quantum critical points.
 It is first placed into a superposition state and then coupled to the system undergoing
 the QPT.
 When the two eigenstates become correlated to two different phases,
 the superposition decoheres.
 The loss of coherence is thus a direct measure of the QPT.

 We have applied this procedure to two types of QPTs,
choosing the couplings between the probe and the system in
such a way that the two states of the probe induce slightly
different values of the control parameter \cite{sunPRL06,Paz}.
No details have to be known about the phases on the two sides
 of the phase transition.
 Only one qubit is measured for the detection of the critical points,
independent of the size of the simulated quantum system. Hence this
method scales very favorably with the size of the system
\cite{Somma02,efficent}.
Theoretical results indicate that the overlap $L$ remains a useful measure
for larger systems in Ising and $XY$ spin chains
\cite{sunPRL06}. For the more complex quantum phase transitions
where many states are close to the ground state (e. g. spin glass),
our fidelity method seems to work, although the details are
still being worked out \cite{praviteComm}.

In the present example, the probe qubit was coupled to all system
qubits in a symmetric way. For other systems, the type of coupling
required may depend on the system Hamiltonian and the nature of the
phases on both sides of the QPT. While a full discussion of this
issue is far beyond the scope of this letter, we expect that if the
phase change involves delocalized states (e.g. spin waves), a single
coupling between the probe qubit and one of the system qubits should be
sufficient to detect the phase transition \cite{Rossini}. On the
other hand, if the changes at the QPTs are local, a larger number
of couplings or probe qubits may be required. In the extreme case,
where critical points separate purely local changes, it may be
necessary to couple the probe qubit to every system qubit or to
implement couplings from a single probe qubit to all system qubits.
Even in this worst case scenario, the number of probe qubits (or
operations) only scales linearly with the size of the system; this
should be contrasted to the readout by quantum state tomography,
where the number of measurements increases exponentially with the
system size. In future work, we plan to apply this type of analysis
to the study of different types of phase transition, including
quantum chaos ~\cite{ponebit}. Furthermore, it should be possible to
use this approach for the characterization of decoherence
~\cite{Paz} and errors that occur during quantum information
processing ~\cite{CoryPRA07}.

  We thank Prof. J.-F. Du for helpful discussions. This work is
supported by the Alexander von Humboldt Foundation, the DFG through
Su 192/19-1, and the Graduiertenkolleg No. 726.


\end{document}